\documentclass[sort&compress,final]{aipproc}\layoutstyle{8x11single}
\usepackage{amsmath}\usepackage{amssymb}\usepackage{bm}\usepackage{booktabs}
\begin{document}\title{Beauty Vector Meson Decay Constants from
QCD Sum Rules} % HEPHY-PUB 940/14
\author{Wolfgang Lucha}{address={Institute for High Energy Physics,
Austrian Academy of Sciences, Nikolsdorfergasse 18, A-1050 Vienna,
Austria}}\author{Dmitri Melikhov}{address={Institute for High
Energy Physics, Austrian Academy of Sciences, Nikolsdorfergasse
18, A-1050 Vienna, Austria},altaddress={D.~V.~Skobeltsyn Institute
of Nuclear Physics, M.~V.~Lomonosov Moscow State University,
119991, Moscow, Russia}}\author{Silvano Simula}{address={Istituto
Nazionale di Fisica Nucleare, Sezione di Roma Tre, Via della Vasca
Navale 84, I-00146, Roma, Italy}}\begin{abstract}We present the
outcomes of a very recent investigation of the decay constants of
nonstrange and strange heavy-light beauty vector mesons, with
special emphasis on the ratio of any such decay constant to the
decay constant of the corresponding pseudoscalar meson, by means
of Borel-transformed QCD sum rules. Our results suggest that both
these ratios are below unity.\end{abstract}
\maketitle

\section{Introduction: Decay Constants of Vector vs.~Pseudoscalar
Mesons}Recently, we started to apply the QCD sum-rule approach
\cite{SVZ}, modified by implementation of an improved algorithm of
extracting hadron properties \cite{LMSETa,LMSETb, LMSETc,LMSETd},
to heavy-quark mesons. We analyzed the decay constants of the
charmed~and~beauty pseudoscalar mesons $D,D_s,B,B_s$
\cite{LMSDa,LMSDb}, the bottom-quark mass value emerging from the
beauty-meson decay constants $f_B,f_{B_s}$ \cite{LMS_mb}, and the
decay constants of the charmed vector mesons $D^*,D_s^*$
\cite{LMSD*}. The logically next step is the investigation of the
decay constants $f_{B^*},f_{B^*_s}$ of the vector beauty mesons
$B^*,B_s^*$ \cite{LMSB*}. In this context, of preeminent interest
is whether the decay constant of $B^*$ proves to be larger or
smaller than that of $B$ (and the one of $B_s^*$ larger or smaller
than that~of~$B_s$).

\section{Relating Hadronic Features and Parameters of Underlying
Theory}\emph{QCD sum rules\/} are relations between observable
properties of hadrons and the parameters of QCD --- the quantum
field theory governing strong interactions ---, found by
evaluating correlation functions of time-ordered nonlocal products
of operators interpolating the hadron under consideration
simultaneously at the hadronic level, by insertion of a complete
set of states, and at the QCD level, by means of Wilson's
\emph{operator product expansion\/} (OPE). For a
\emph{pseudoscalar\/} beauty meson $B$ of mass $M_B$ composed of a
bottom quark $b$ of mass $m_b$ and a light quark $q$ of mass
$m_q,$ let's interpolate $B$ by the pseudoscalar current
$j_5(x)=(m_b+m_q)\,\bar q(x)\,{\rm i}\,\gamma_5\,b(x)$ defining
the $B$-meson's decay constant $f_B$ by
$\langle0|j_5(0)|B\rangle=f_B\,M_B^2$:$$\Pi(p^2)\equiv{\rm
i}\int{\rm d}^4x\,{\rm e}^{{\rm i}\,p\,x}\left\langle0\left|
T\!\left(j_5(x)\,j^\dag_5(0)\right)\right|0 \right\rangle.$$The
resulting correlator $\Pi$ can be expressed in terms of a
perturbative contribution of the form of a dispersion integral of
a spectral density $\rho_{\rm pert}$ and a non-perturbative (NP)
or ``power'' contribution involving so-called ``vacuum
condensates''. A Borel transformation to the Borel variable $\tau$
removes subtraction terms and suppresses higher hadronic
contributions:$$\Pi(\tau)=f_B^2\,M_B^4\exp\left(-M_B^2\,\tau\right)+
\int\limits_{(M_{B^*}+M_P)^2}^\infty{\rm d}s\,{\rm e}^{-s\,\tau}\,
\rho_{\rm hadr}(s)=\int\limits_{(m_b+m_q)^2}^\infty{\rm d}s\,{\rm
e}^{-s\,\tau}\,\rho_{\rm pert}(s,\mu)+\Pi_{\rm power}(\tau,\mu)\
.$$The lower limit of the integral of the hadron spectral density
$\rho_{\rm hadr}$ involves the mass $M_P$ of the lightest
relevant~meson $P.$ Assuming these higher hadronic contributions
to compensate all the perturbative QCD contributions above the
\emph{effective threshold\/} $s_{\rm eff}(\tau)$ results in the
sum rule sought, where we take the liberty to label its QCD side
the dual~correlator $\Pi_{\rm dual}$:
\begin{equation}f_B^2\,M_B^4\exp\left(-M_B^2\,\tau\right)
=\int\limits_{(m_b+m_q)^2}^{s_{\rm eff}(\tau)}{\rm d}s\,{\rm
e}^{-s\,\tau}\,\rho_{\rm pert}(s,\mu)+\Pi_{\rm
power}(\tau,\mu)\equiv\Pi_{\rm dual}(\tau,s_{\rm eff}(\tau))\
.\label{Eq:SR}\end{equation}The perturbative-QCD \emph{spectral
density\/}, $\rho_{\rm pert},$ in form of its expansion in powers
of the strong coupling $\alpha_{\rm s}(\mu)$ in the $\overline{\rm
MS}$ renormalization scheme involving the $b$-quark's pole mass
$M_b,$ is presently available up to three-loop
accuracy~\cite{SDa,SDb}:$$\rho_{\rm
pert}(s,\mu)=\rho_0(s,M_b^2)+\frac{\alpha_{\rm s}(\mu)}{\pi}\,
\rho_1(s,M_b^2)+\frac{\alpha_{\rm s}^2(\mu)}{\pi^2}\,
\rho_2(s,M_b^2,\mu)+\cdots\ .$$Conversion to the $b$-quark's mass
$\overline{m}_b(\nu)$ defined in the $\overline{\rm MS}$
renormalization scheme is effected, for known $r_1,r_2$
\cite{JL},~by$$M_b=\overline{m}_b(\nu)\left(1+\frac{\alpha_{\rm
s}(\nu)}{\pi}\,r_1+\frac{\alpha_{\rm
s}^2(\nu)}{\pi^2}\,r_2+\cdots\right).$$For generality, we allow
the renormalization scales of the strong coupling, $\mu,$ and of
the $b$-quark mass, $\nu,$ to be unequal.

Upon adopting as starting point of such QCD sum-rule analysis the
interpolating \emph{vector\/} current $j_\mu(x)\equiv\bar
q(x)\,\gamma_\mu\,b(x),$ a (\emph{mutatis mutandis\/}) similar
formalism may be established for studying the features of the
beauty vector mesons $B^*,B_s^*.$

\section{Advanced Extraction of Hadron Features, such as Decay
Constants}In the course of applying QCD sum rules, two steps
demand particular consideration: the reasonable choice of the
Borel variable and the determination of the behaviour of the
effective continuum threshold as a function of this
Borel~variable.\begin{description}\item[Borel working window.] The
optimization of the extracted QCD sum-rule predictions for hadron
observables provides two constraints that delimit, from both below
and above, the interval of acceptable values of the Borel
parameter~$\tau$: the relative contribution of the ground state
(which increases with increasing value of $\tau$) to the
correlator should be sufficiently large and the relative magnitude
of the power corrections should remain under control
(which~becomes increasingly difficult for rising value of $\tau$).
For, \emph{e.g.}, the pseudoscalar mesons $B$ and $B_s,$ this
confines $\tau$ to the~range$$0.05\;\mbox{GeV}^{-2}\lessapprox
\tau\lessapprox0.175\;\mbox{GeV}^{-2}\ .$$\item[Effective
continuum threshold.] The effective threshold $s_{\rm eff}(\tau)$
serves to encode the way all higher contributions~from
perturbative QCD are dealt with by assuming an exact cancellation
against the hadronic excitations and continuum and thus
constitutes, in addition to the perturbative spectral density and
the power corrections, a \emph{crucial ingredient\/} of the QCD
sum-rule method. In order to formulate a criterion for finding the
function $s_{\rm eff}(\tau),$ we introduce,~for the example of the
$B$ meson, the ($\tau$-dependent) \emph{dual\/} mass $M_{\rm
dual}(\tau)$ and \emph{dual\/} decay constant $f_{\rm dual}(\tau)$
by the~definitions, obviously inspired by the particular way these
two hadron characteristics enter in the Borelized QCD sum
rule~(\ref{Eq:SR}),$$M_{\rm dual}^2(\tau)\equiv-\frac{{\rm
d}}{{\rm d}\tau}\log\Pi_{\rm dual}(\tau,s_{\rm eff}(\tau))\
,\qquad f_{\rm
dual}^2(\tau)\equiv\frac{\exp\!\left(M_B^2\,\tau\right)}{M_B^4}\,
\Pi_{\rm dual}(\tau,s_{\rm eff}(\tau))\ .$$In order to infer the
dependence on $\tau$ of the effective threshold from the knowledge
of the meson mass $M_B,$ we start from an ansatz for $s_{\rm
eff}(\tau),$ for which (because of the limited $\tau$ range) it is
sufficient to use a polynomial~of~degree~$n,$\begin{equation}
s^{(n)}_{\rm eff}(\tau)=\sum_{j=0}^ns^{(n)}_j\,\tau^j\ ,\qquad
n=0,1,2,\dots\ ,\label{Eq:ETA}\end{equation}and determine the
expansion coefficients, $s^{(n)}_j,$ by minimizing the squared
differences of the squares of \emph{dual\/}~meson mass and
\emph{measured\/} meson mass over a set of $N$ equidistant points
$\tau_i$ in the adopted Borel interval, \emph{i.e.},~the~quantity
$$\chi^2\equiv\frac{1}{N} \sum_{i=1}^N\left[M^2_{\rm
dual}(\tau_i)-M_B^2\right]^2\ ,\qquad N=1,2,3,\dots\
.$$\end{description}

\subsection{Uncertainties of decay constants from the advanced
extraction algorithm for QCD sum rules}All OPE parameters and the
limited accuracy of the QCD sum-rule method add to the
uncertainties of its predictions.\begin{description}
\item[OPE-related error.] Assuming Gaussian distributions for all
parameters entering in the OPE except for the scales $\mu,\nu,$
for which we assume uniform distributions in the range
$3\;\mbox{GeV}\leq\mu,\nu\leq6\;\mbox{GeV},$ bootstrapping yields
a~distribution of decay constants close to a Gaussian shape. Thus,
we feel save if regarding the OPE-related error to be Gaussian.
\item[Systematic error.] We estimate this intrinsic error from the
spread of results from linear through cubic ansatzes $s^{(n)}_{\rm
eff}(\tau).$\end{description}

\section{Decay Constants of Pseudoscalar ($\bm{f}_{\!\bm{B}}$) vs.
Vector Beauty Mesons ($\bm{f}_{\!\bm{B}^{\bm{*}}}$)}Having pinned
down the variationally optimal effective-threshold polynomial
$s^{(n)}_{\rm eff}(\tau)$ for fixed $n,$ it is straightforward to
extract the QCD sum-rule prediction for the dual decay constants
of $B,B_s,B^*,B_s^*.$ As OPE input, we use the more or less
standard numerical values for $\overline{\rm MS}$-scheme quark
masses, strong coupling, and lowest-dimensional vacuum~condensates
\begin{align*}&\overline{m}_d(2\;\mbox{GeV})=(3.5\pm0.5)\;\mbox{MeV}\
,\qquad\overline{m}_s(2\;\mbox{GeV})=(95\pm5)\;\mbox{MeV}\
,\qquad\overline{m}_b(\overline{m}_b)=(4.18\pm0.03)\;\mbox{GeV}\
,\\&\alpha_{\rm s}(M_Z)=0.1184\pm0.0007\ ,\qquad\langle\bar
qq\rangle(2\;\mbox{GeV})=-[(269\pm17)\;\mbox{MeV}]^3\
,\\&\langle\bar ss\rangle(2\;\mbox{GeV})=(0.8\pm0.3)\,\langle\bar
qq\rangle(2\;\mbox{GeV})\ ,\qquad\left\langle\frac{\alpha_{\rm
s}}{\pi}\,GG\right\rangle =(0.024\pm0.012)\;\mbox{GeV}^4\
.\end{align*}

Optimization of our results requires careful choices of both
renormalization scheme (defining the heavy-quark mass) and
renormalization scale. In principle, QCD sum-rule results for
observables should not depend on such technicalities. Truncations
of perturbative expansions to finite powers of $\alpha_{\rm s}$
and of power corrections to vacuum condensates of lowest
dimensions, however, induce a clearly unphysical dependence on
renormalization details. Confidence in our findings is, of course,
strengthened if we can arrange for a distinct hierarchy of the
various contributions to the correlator. However, unlike the
charmed-meson case, where we found only a mild
renormalization-scale sensitivity, the beauty-meson decay
constants and, in particular, their ratios
$f_{B^*_{(s)}}/f_{B_{(s)}}$ are the result of a delicate
\cite{LMSB@} interplay of the different~contributions.

\vspace{-0.343ex}
\subsection{First point of reference: Dual decay constant
$\bm{f}_{\!\bm{B}}$ of pseudoscalar nonstrange beauty meson
$\bm{B}$}
\begin{figure}[hb]\begin{tabular}{cc}\label{Fig:fB}
\includegraphics[width=.491\columnwidth]{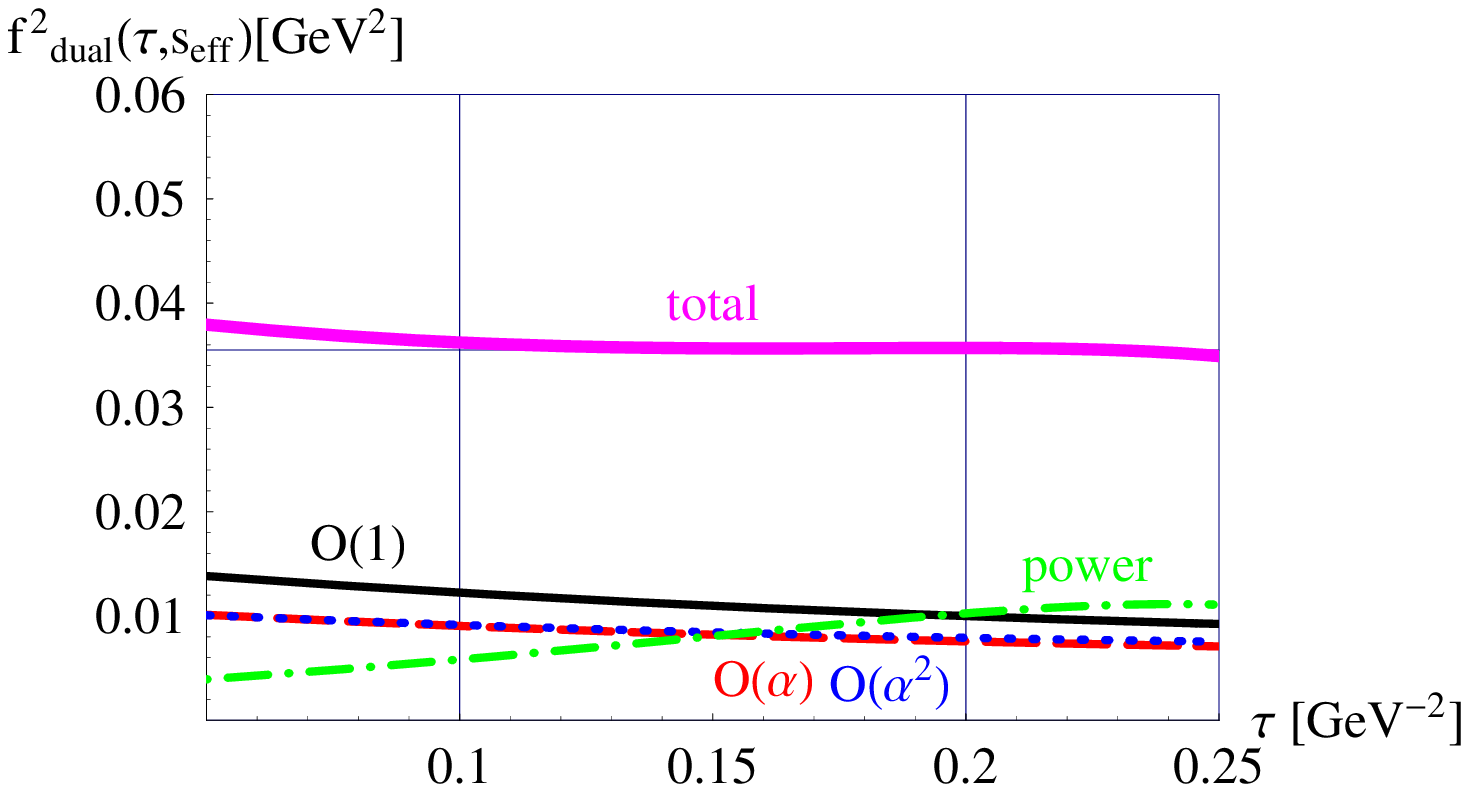}&
\includegraphics[width=.491\columnwidth]{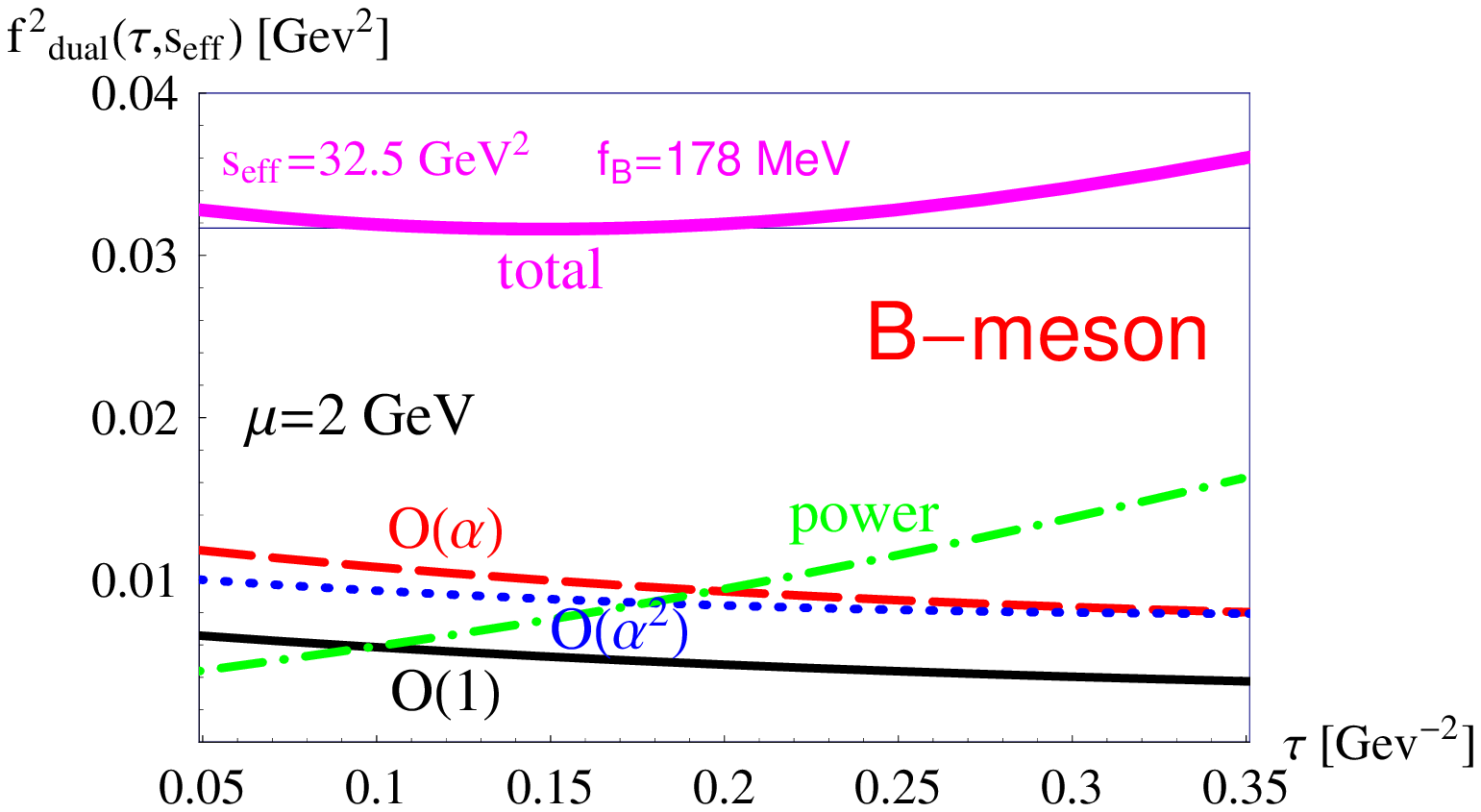}\\(a)&(b)\\
\includegraphics[width=.491\columnwidth]{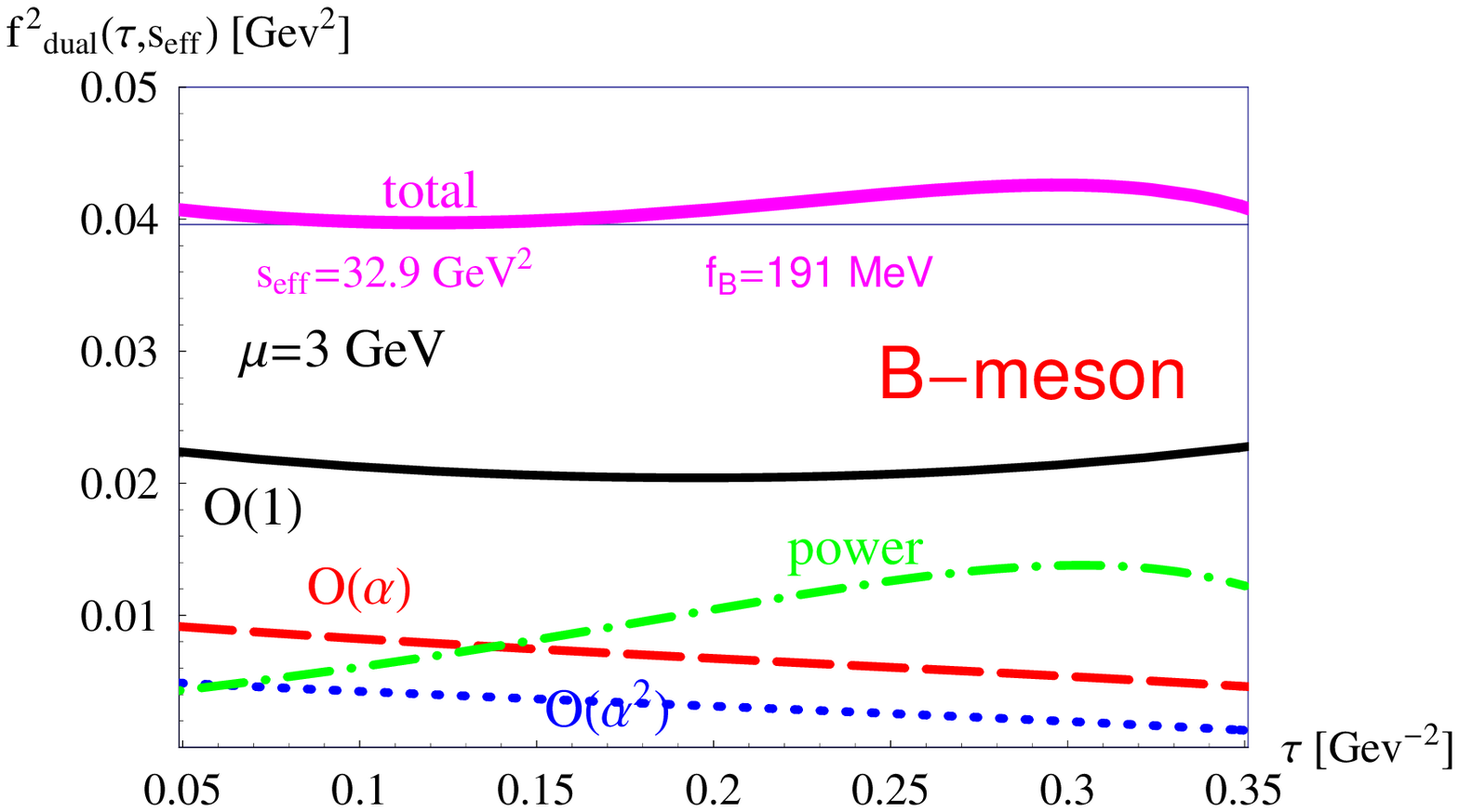}&
\includegraphics[width=.491\columnwidth]{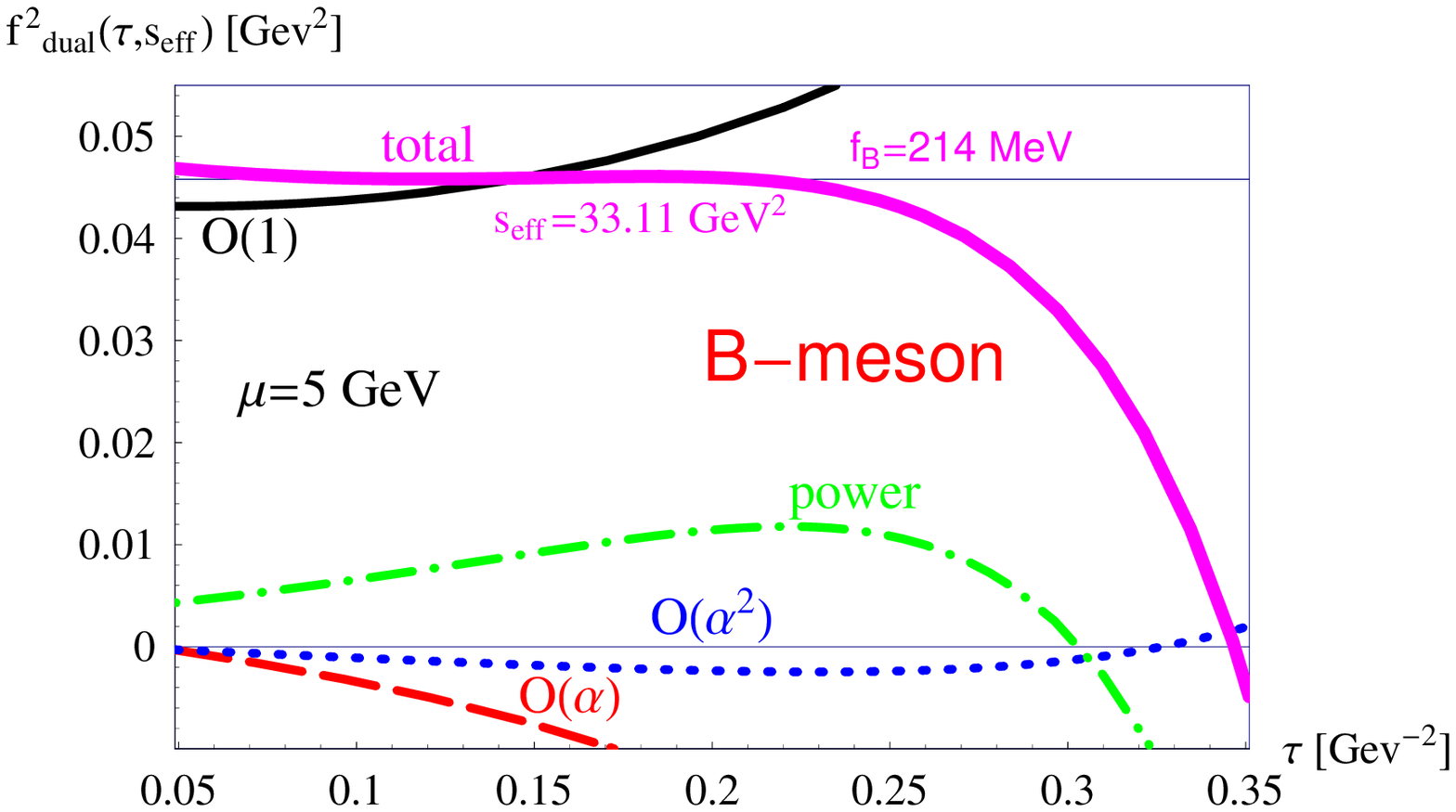}\\(c)&(d)
\caption{Dual decay constant $f_{\rm dual}(\tau)$ of the $B$
meson: breakdown of the OPE contributions arising for fixed
effective threshold $s_{\rm eff}$ and either $b$-quark pole mass
(a) or $b$-quark $\overline{\rm MS}$ mass at renormalization
scales $\mu=2\;\mbox{GeV}$ (b), $\mu=3\;\mbox{GeV}$ (c), and
$\mu=5\;\mbox{GeV}$(d).}\end{tabular}\end{figure}

Figure \ref{Fig:fB}, serving as solid basis of
comparison, summarizes our $B$-meson decay-constant results
arising for a constant effective threshold: relying on the
$b$-quark $\overline{\rm MS}$ instead of pole mass and a
sophisticated choice of $\mu$ is clearly preferable.

\subsection{Actual primary target of interest: Decay constant
$\bm{f}_{\!\bm{B}^{\bm{*}}}$ of vector nonstrange beauty meson
$\bm{B}^{\bm{*}}$}
\begin{figure}[hb]\begin{tabular}{cc}\label{Fig:fB*}
\includegraphics[width=.491\columnwidth]{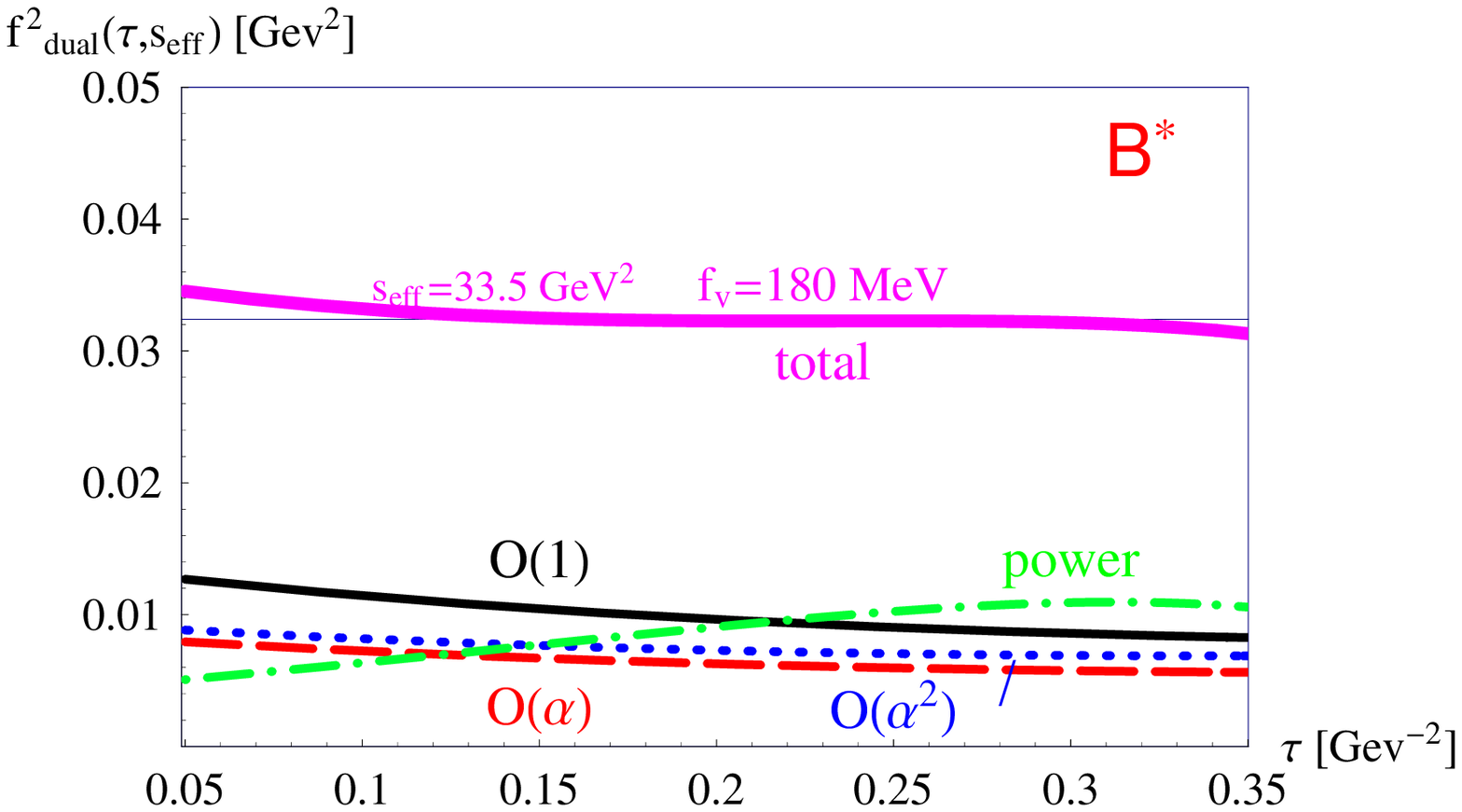}&
\includegraphics[width=.491\columnwidth]{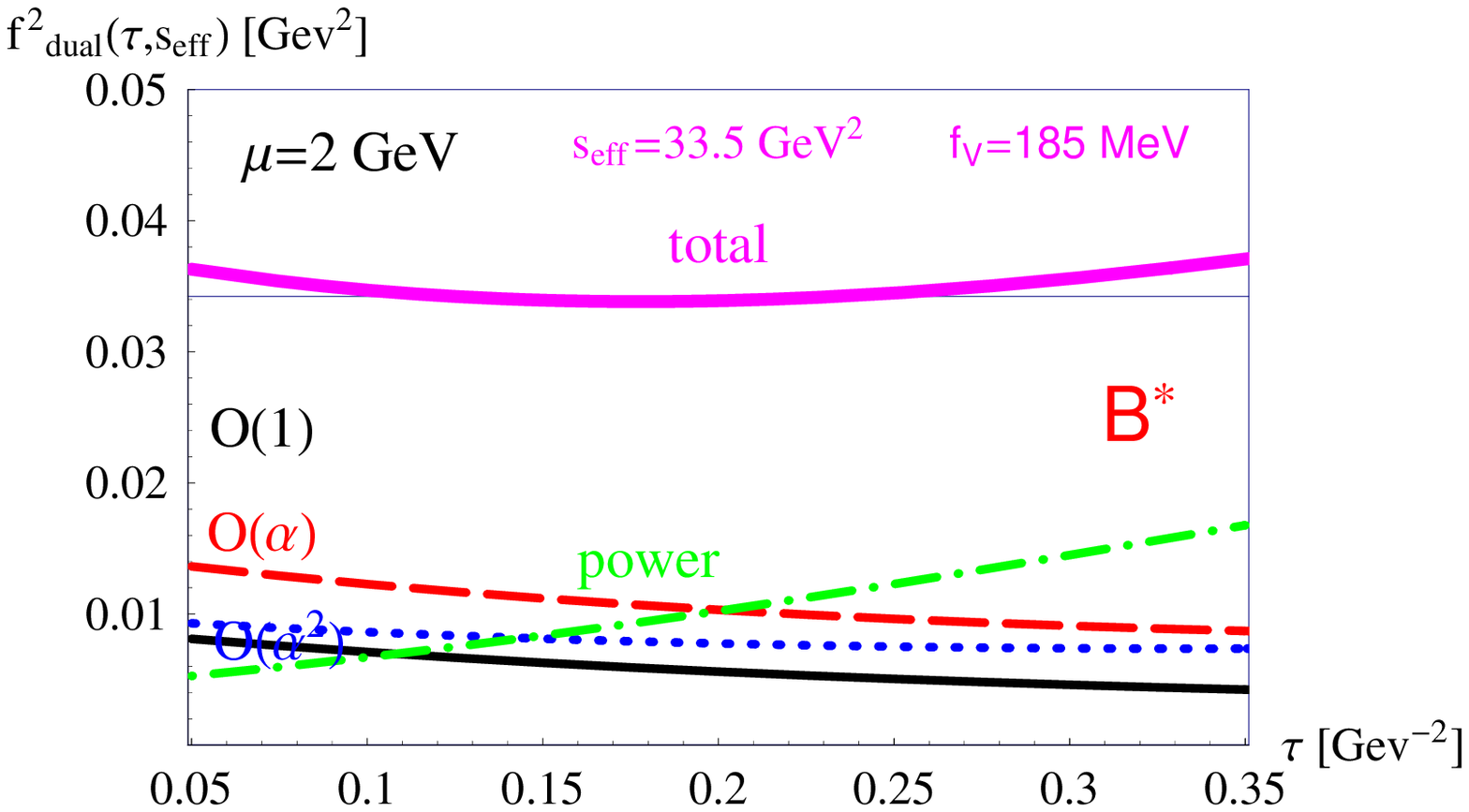}\\(a)&(b)\\
\includegraphics[width=.491\columnwidth]{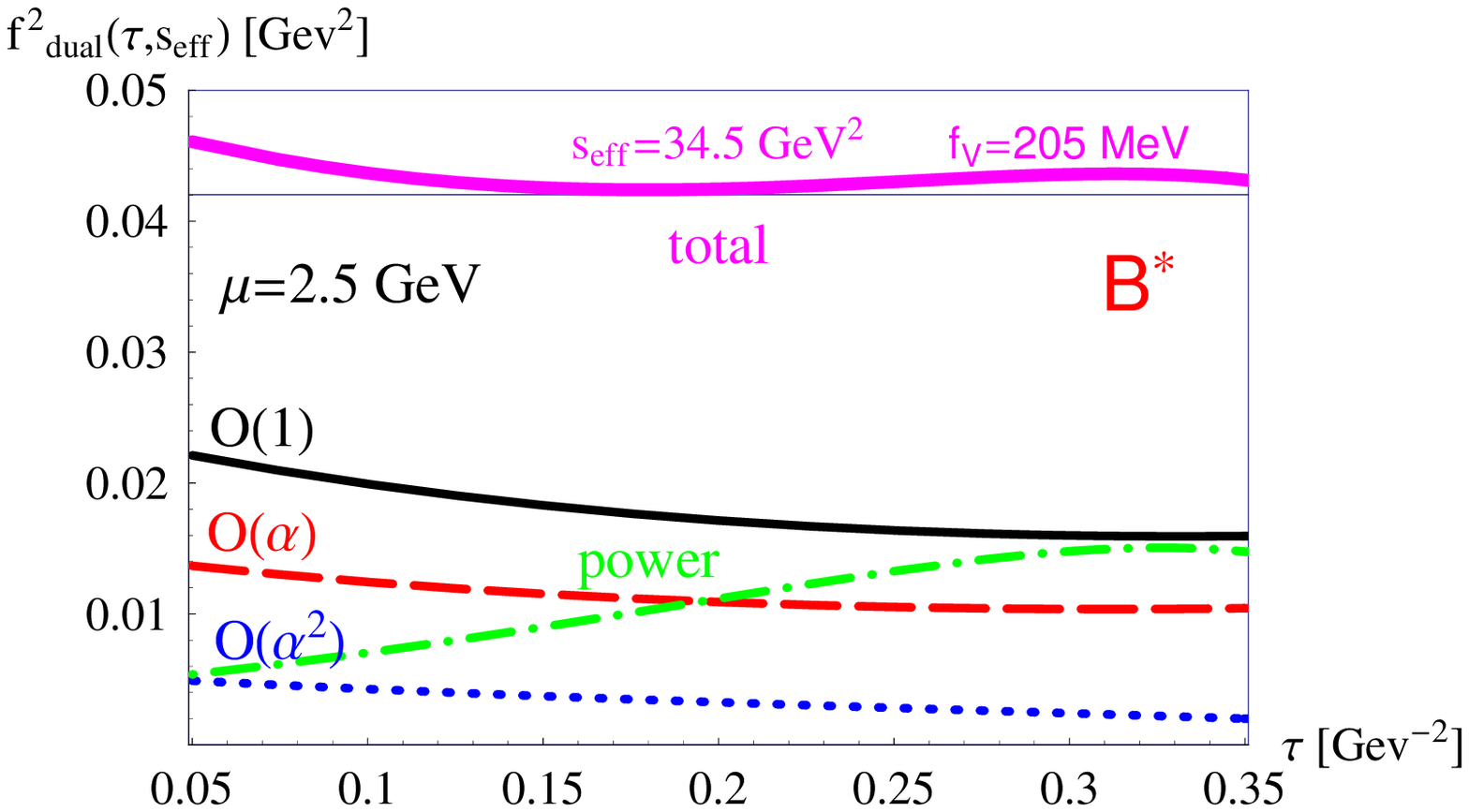}&
\includegraphics[width=.491\columnwidth]{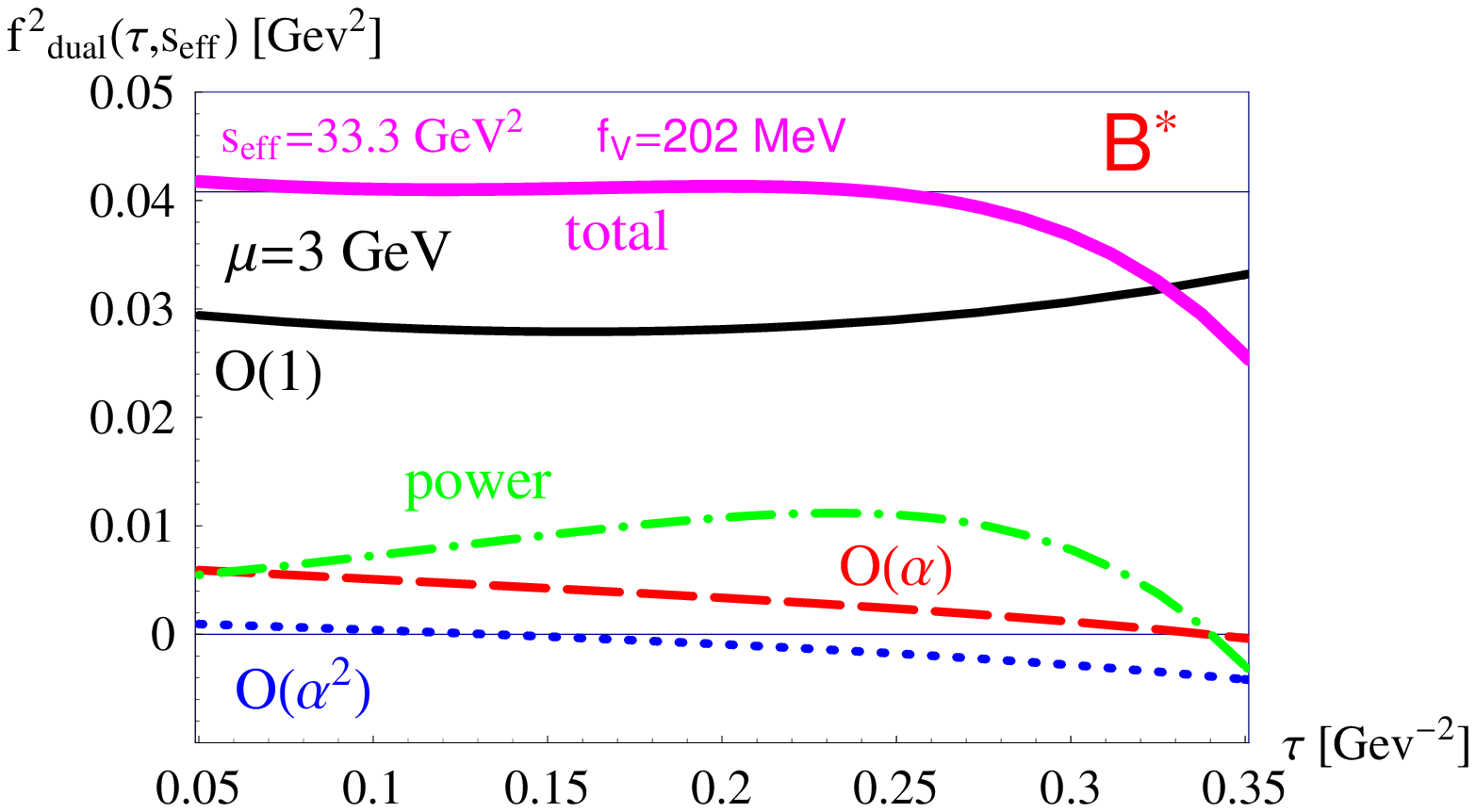}\\(c)&(d)\\
\includegraphics[width=.491\columnwidth]{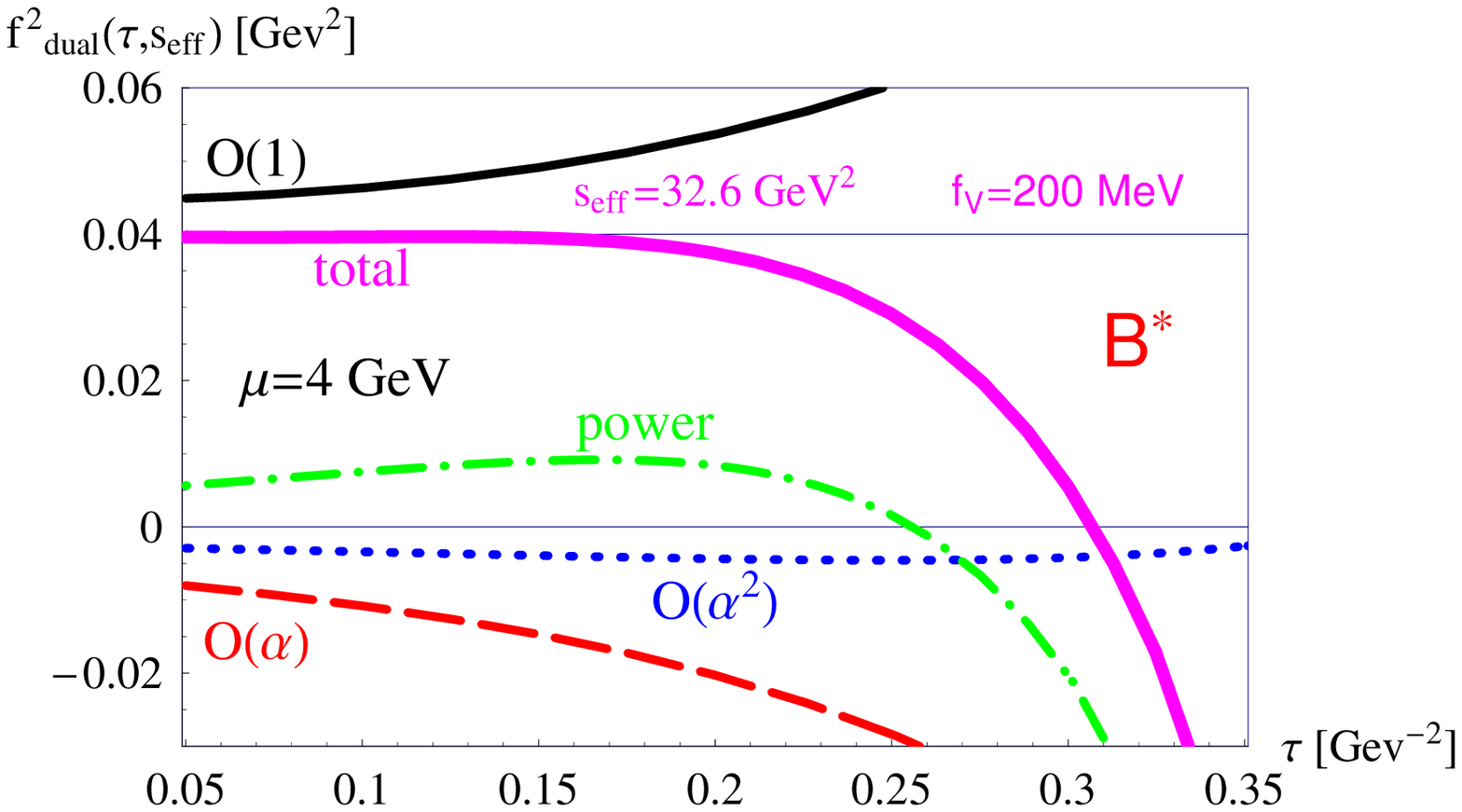}&
\includegraphics[width=.491\columnwidth]{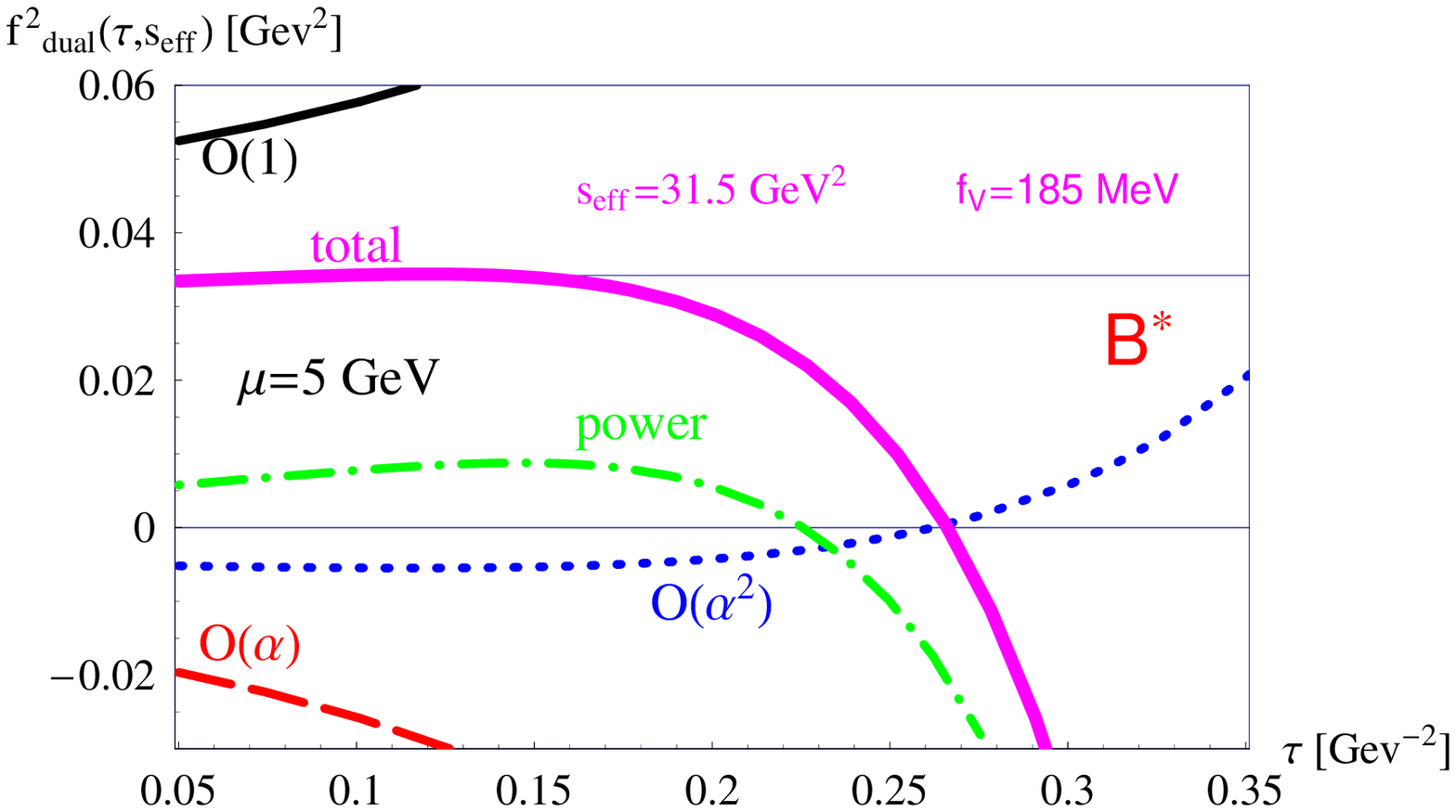}\\(e)&(f)
\caption{Dual decay constant $f_{\rm dual}(\tau)$ of the $B^*$
meson: breakdown of the OPE contributions, found if adopting a
fixed~effective threshold $s_{\rm eff}$ and either the pole-mass
(a) or the $\overline{\rm MS}$-mass definition of the $b$-quark's
mass, for several choices of the renormalization scale in the
interval $2\;\mbox{GeV}\le\mu\le5\;\mbox{GeV},$ namely,
$\mu=2\;\mbox{GeV}$ (b), $\mu=2.5\;\mbox{GeV}$ (c),
$\mu=3\;\mbox{GeV}$ (d), $\mu=4\;\mbox{GeV}$ (e) and
$\mu=5\;\mbox{GeV}$~(f).}\end{tabular}\end{figure}

Figure \ref{Fig:fB*} illustrates the impact of
definition of the heavy-quark mass and choice of renormalization
scale on the decay constant of the beauty vector meson $B^*$
resulting for $s_{\rm eff}(\tau)=\mbox{const}$ and the $b$-quark
$\overline{\rm MS}$ mass
$\overline{m}_b(\overline{m}_b)=4.18\;\mbox{GeV}$~\cite{PDG}:
\begin{itemize}\item Using the $b$-quark's pole mass, we obtain no
perturbative hierarchy. All contributions are of roughly the
same~size.\item In the $\overline{\rm MS}$ scheme, perturbative
convergence depends on $\mu$ and may be achieved by adopting a
sufficiently large~$\mu.$\item All plots shown in
Fig.~\ref{Fig:fB*} exhibit, for the extracted decay constants,
approximately flat plateaus over wide ranges of $\tau$. This
observation tells us (anew) that the requirement of \emph{Borel
stability\/} does not necessarily yield reliable~results.
\end{itemize}

\section{Dependence of the Decay Constants on the Renormalization
Scale}
\begin{figure}[hb]\begin{tabular}{cc}\label{Fig:ratio}
\includegraphics[width=.491\columnwidth]{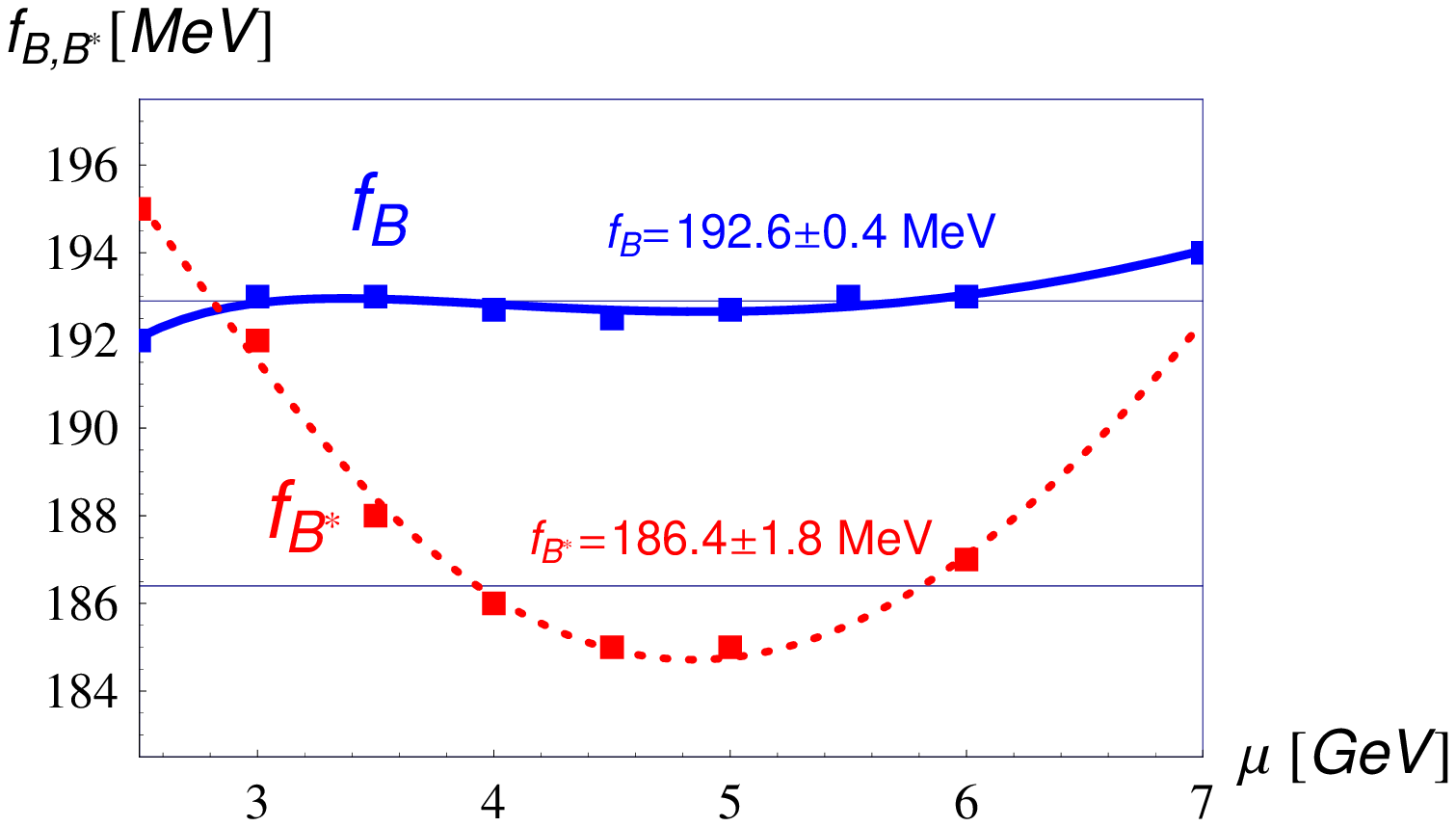}&
\includegraphics[width=.491\columnwidth]{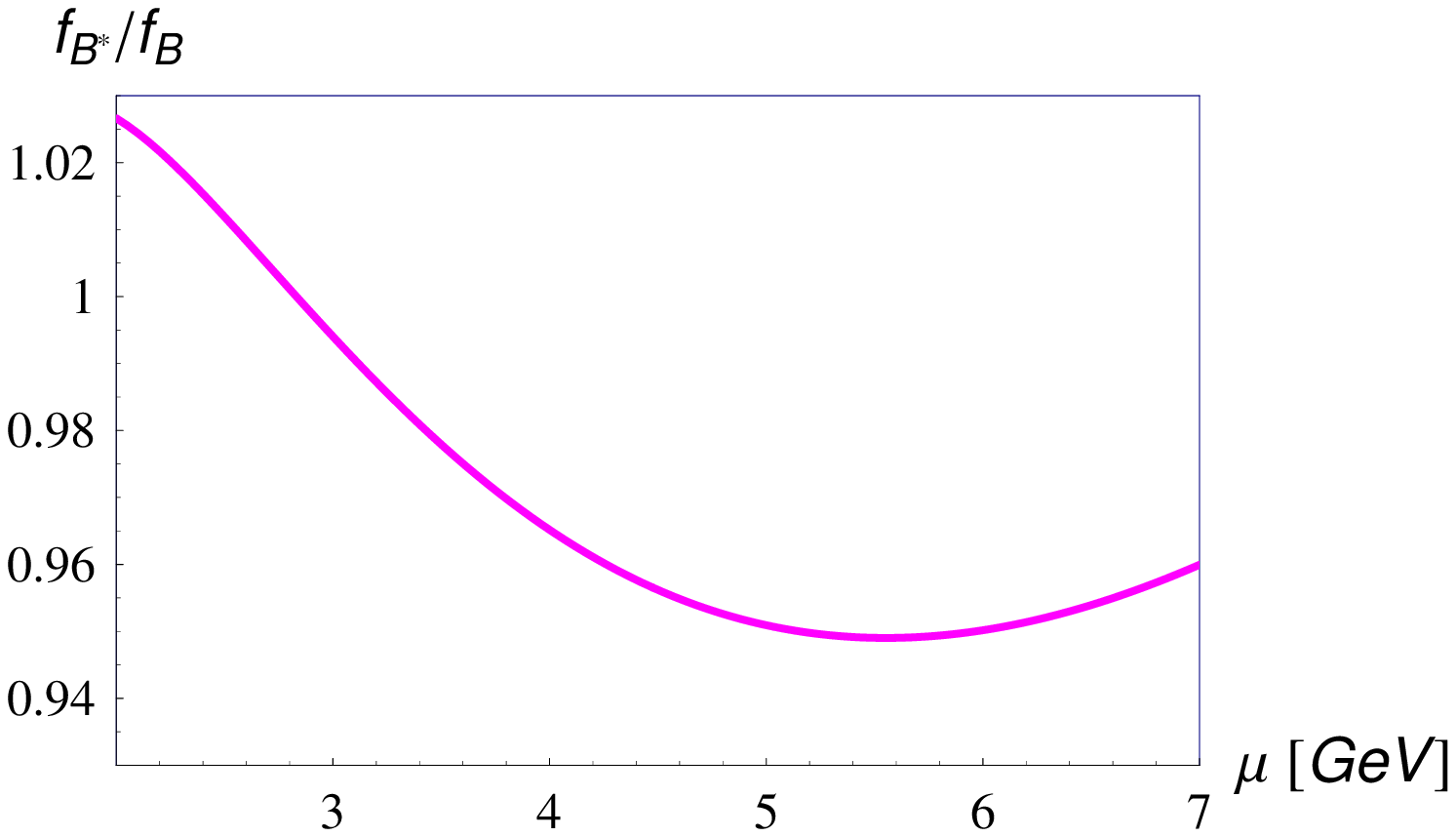}
\\(a)&(b)\caption{Dependence of our results for the decay constants
$f_B,f_{B^*}$ (a) and their ratio $f_{B^*}/f_B$ (b) on the
renormalization scale~$\mu.$}\end{tabular}\end{figure}

The strong sensitivity of $f_B$ to the $b$-quark mass
allowed us to deduce the latter \cite{LMS_mb} from the lattice-QCD
findings~for~$f_B$:$$\overline{m}_b(\overline{m}_b)
=(4.247\pm0.027_{\rm OPE}\pm0.018_{\rm exp}\pm0.011_{\rm
syst})\;\mbox{GeV}=(4.247\pm0.034)\;\mbox{GeV}\ .$$The behaviour
of our numerical results for the dual decay constants of the
nonstrange beauty mesons as functions~of the renormalization scale
$\mu$ in the proximity of an average $\overline{\mu}$ of $\mu$
defined by requiring $f_{\rm dual}(\overline{\mu})=\langle f_{\rm
dual}(\mu)\rangle$ (cf.~Fig.~\ref{Fig:ratio}) and as functions of
the $\overline{\rm MS}$ $b$-quark mass
$m_b\equiv\overline{m}_b(\overline{m}_b)$ may be
\emph{approximately\/} reproduced by a linear decrease with~$m_b$
and a series expansion in powers of the logarithm of $\mu.$ For
the pseudoscalar meson $B,$ the $\mu$ dependence is
not~very~dramatic:\begin{align*}&f_B^{\rm
dual}(m_b,\mu=\overline{\mu})=192.6\;\mbox{MeV}-13\;\mbox{MeV}
\left(\frac{m_b-4.247\;\mbox{GeV}}{0.034\;\mbox{GeV}}\right),
\qquad\overline{\mu}=5.59\;\mbox{GeV}\ ,\\&f_B^{\rm
dual}(m_b=4.247\;\mbox{GeV},\mu)=192.6\;\mbox{MeV}
\left(1-0.0015\log\frac{\mu}{\overline{\mu}}
+0.030\log^2\frac{\mu}{\overline{\mu}}
+0.061\log^3\frac{\mu}{\overline{\mu}}\right).\end{align*}For the
vector meson $B^*,$ the coefficients are markedly larger, that of
the linear term by almost two orders of magnitude:\begin{align*}
&f_{B^*}^{\rm
dual}(m_b,\mu=\overline{\mu})=186.4\;\mbox{MeV}-10\;\mbox{MeV}
\left(\frac{m_b-4.247\;\mbox{GeV}}{0.034\;\mbox{GeV}}\right),
\qquad\overline{\mu}=5.82\;\mbox{GeV}\ ,\\&f_{B^*}^{\rm
dual}(m_b=4.247\;\mbox{GeV},\mu)=186.4\;\mbox{MeV}
\left(1+0.106\log\frac{\mu}{\overline{\mu}}
+0.337\log^2\frac{\mu}{\overline{\mu}}
+0.173\log^3\frac{\mu}{\overline{\mu}}\right).\end{align*}The very
pronounced dependence of the $B^*$ decay constant on the
renormalization scale $\mu$ has, of course, a considerable impact
on the ratio $f_{B^*}/f_B$ of beauty-meson vector-to-pseudoscalar
decay constants: this ratio proves to emerge slightly above unity
for small $\mu$ (say, $\mu\lessapprox3\;\mbox{GeV}$) but certainly
below unity for larger $\mu$ (\emph{i.e.},
$\mu\gtrapprox3\;\mbox{GeV}$), as evident
from~Fig.~\ref{Fig:ratio}.

\section{Condensed Summary of Outcomes, Discussion, Conclusion,
Outlook}The extraction of the decay constants of heavy-light
mesons from Borelized QCD sum rules forms in the beauty-meson
sector, for, at least, two reasons, a definitely greater challenge
than it does in the charmed-meson sector. First,~for~beauty
mesons, we encounter a pronounced dependence of our predictions on
the mass of the $b$ quark. Secondly, for the beauty mesons, the
predictions develop an unpleasantly strong dependence on the
adopted renormalization scale $\mu.$ Inspecting the behaviour of
the decay constants under variation of $\mu,$ we find that the
decay constant of the \emph{pseudoscalar\/} meson $B,$ $f_B,$
increases with rising $\mu$ (see Fig.~\ref{Fig:fB}), whereas the
one of the \emph{vector\/} meson $B^*,$ $f_{B^*},$ decreases with
rising $\mu$ (cf.~Fig.~\ref{Fig:fB*}). Averaging our findings for
a quadratic effective-threshold polynomial (\ref{Eq:ETA}) over the
interval $3\;\mbox{GeV}\leq\mu\leq6\;\mbox{GeV},$ we~find
\begin{align*}f_B&=(192.6\pm1.6)\;\mbox{MeV}\ ,&\quad
f_{B^*}&=(186.4\pm3.2)\;\mbox{MeV}\ ,\\f_{B_s}&=(231.0\pm
1.8)\;\mbox{MeV}\ ,&\quad f_{B_s^*}&=(215.2\pm3.0)\;\mbox{MeV}\
,\end{align*}where the errors take into account only the effects
of the variation of $\mu.$ The resulting decay-constant ratios
(Fig.~\ref{Fig:ratio})~read$$\frac{f_{B^*}}{f_B}=0.923\pm0.059\
,\qquad \frac{f_{B_s^*}}{f_{B_s}}=0.932\pm0.047\ .$$Surprisingly,
the QCD sum-rule prediction for $f_{B^*}/f_B$ is below the
corresponding results from lattice QCD, which~seem to favour a
value slightly above but still close to unity \cite{LMS-L}.
Clearly, such tension calls for further detailed investigations.

\bibliographystyle{aipproc}\end{document}